\documentclass[apjl]{emulateapj}
\usepackage{mathptmx}

\begin{document}

\title{Mass Loss by X-ray Winds from Active Galactic Nuclei}
\author{Doron Chelouche\altaffilmark{1}}
\altaffiltext{1} {Canadian Institute for Theoretical Astrophysics, University of Toronto, 60 St. George st., Toronto ON M5S 3H8, Canada; doron@cita.utoronto.ca}
\shortauthors{Chelouche D.}
\shorttitle{Mass Loss by X-ray Winds from AGN}

\begin{abstract}

We consider a sample of type-I active galactic nuclei (AGN) that were observed by {\it Chandra/HETG} and resulted in high signal-to-noise grating spectra, which we study in detail. All
objects show signatures for very high ionization outflows. Using a novel scheme to model the physics and spectral signatures of gaseous winds from these objects, we are able to estimate the mass loss rates and kinetic luminosities associated
with the highly ionized gas and investigate its physical properties. Our conclusions are as follows: 1) There is
a strong indication that the outflowing gas in those objects is
multi-phase with similar kinematics for the different phases.
2) The X-ray spectrum is consistent with such flows being
thermally driven from $\sim$\,pc scales, and are therefore unlikely
to be associated with the inner accretion disk. 3) The underlying
X-ray spectrum consists of a hard X-ray powerlaw which is
similar for all objects shining below their Eddington rate and
a soft excess whose contribution becomes more prominent for
objects shining close to their Eddington limit. 4) The physical properties of the outflow are similar in all cases and a coherent picture emerges concerning its
physical properties. 5) The deduced mass loss rates are, roughly, of the order of the mass accretion rate in those objects so that the kinetic luminosity carried by such winds is only a tiny fraction ($\ll$1\%) of the bolometric luminosity. We discuss the implications of our results for AGN structure and AGN interaction with the environment.
\end{abstract}

\keywords{ acceleration of particles --- ISM: jets and outflows
--- galaxies: Seyfert --- quasars: absorption lines --- X-rays:
individual (NGC\,3783, NGC\,5548, NGC\,7469, NGC\,4151, MCG-6-30-15)}

\section{Introduction}

Recent {\it Chandra} and {\it XMM} grating observations of type-I
(broad emission lines) active galactic nuclei (AGN) show the
presence of highly ionized gas (HIG) outflowing from the centers
of many objects (e.g., Kaastra et al. 2000, Kaspi et al. 2001,
Pounds et al. 2003, McKernan et al. 2007). This suggests that HIG may play an important
role in AGN physics and so must be included in future unification schemes
(e.g., Elvis 2000). Currently, however, little is known with
confidence about the physical properties of this component and a
clear physical picture is yet to emerge. In addition to being
important for AGN study, HIG flows carry energy and momentum out
from their inner engine to potentially large scales and, as such, may have a profound
impact on the environment of quasars, be it their host galaxies or even the
inter-galactic medium (e.g., Di Matteo et al. 2005, Scannapieco \& Oh
2003). Thus, constraining the physical properties of AGN outflows is a fundamental question in AGN
physics with great implications for a wide range of astrophysical
problems.

Detailed X-ray spectroscopic observations of a few nearby low-luminosity AGN (Seyfert
1 galaxies) indicate that HIG outflows have a stratified ionization
structure and that they are located within a parsec or so from the
central continuum source (e.g., Blustin et al. 2007,  Kaspi \& Behar 2006, Kraemer et al.
2005, Krongold et al. 2003, 2009, Netzer et al. 2003, Schulz et al. 2008, Smith et al. 2007, Steenbrugge et al. 2005) . The 
outflowing gas appears to be multiphase with various phases being
in a rough pressure equilibrium and lying on the thermally stable 
parts of the heating-cooling curve (e.g., Chakravorty et al. 2008, Holczer et al. 2007, 
Netzer et al. 2003). Despite the considerable progress in our
phenomenological understanding of HIG flows, their underlying physics
remains elusive. For example, a detailed investigation of the best studied HIG in 
AGN (the case of NGC\,3783) yielded  mass loss rates in the range of a few percent to 
nearly a hundred solar masses per year (e.g., Behar et al. 2003, Blustin et al. 2005, Netzer et al. 2003). 
Some studies pre-suppose that the mass loss rate is less than or equal to the mass accretion rate (e.g.,
Steenbrugge et al. 2005) yet it is not clear that this must be the case. Clearly, the current situation is
unsatisfactory if we wish to understand the role of HIG flows in
the global picture of AGN and assess their effect on the environment of such objects.

Perhaps the main limitation that prevents us from fully understanding the
HIG phenomenon in AGN results from model incompleteness: for example, some
works dealing with HIG absorption measure the column
densities from individual transitions and attempt to estimate the mass loss
rate by requiring full volume filling gas, which does not seem to be supported
by recent observational and theoretical work; e.g., Arav, Li, \&
Begelman 1994, Kraemer et al. 2005, Proga et al. 2008). Other works make use of
detailed photoionization calculations and spectral modeling to
determine the temperature and column density range of the HIG (e.g.,
Kaspi et al. 2002,  Krongold et al. 2003, Netzer et al. 2003, R{\'o}{\.z}a{\'n}ska et al. 2006).
Nevertheless, pure photoionization modelling does not account for
the kinematics of the outflow and cannot be used to reveal the
location of the gas. This is due to an inherent degeneracy between the density of
the photoionized gas and the distance from the ionizing source for a wide 
range of densities relevant to HIG outflows (see however R{\'o}{\.z}a{\'n}ska et al. 2008). Limits on the location of the gas may be obtained by studying the reaction of the photo-ionized gas to the varying ionizing flux level of the source, as has been done for the case of NGC\,3783 (Netzer et al. 2003, Krongold et al. 2005; see also Chevallier et al. 2007). New numerical schemes that self-consistently model the gas dynamics as well as its thermal state were recently calculated by
Dorodnitsyn et al. (2008a,b) and their spectral predictions are yet to be compared to observations. Thus, the physics of HIG outflows in AGN and, in particular, the estimates for the mass carried by them are subject to considerable uncertainties.

Recently, Chelouche \& Netzer (2005; hereafter CN05) have
demonstrated that, by employing a detailed and self-consistent
modelling of the dynamics and photoionization properties of the
outflowing gas, it may be possible to constrain the mass loss rate
to much better precision than before. By applying their model to
NGC\,3783, they found that the mass loss rate is considerably
smaller compared to previous estimates. While their model seems to
be consistent with most observational constraints for the case of
NGC\,3783, it is yet to be confirmed for other AGN. Here we wish to adopt a more general approach and study a sample of X-ray bright, nearby type-I AGN and investigate their properties using a consistent analysis method. Specifically, we wish to test the thermally driven wind model as a possible new ingredient of the AGN structure and do that by considering several objects where such advanced analysis is warranted. Our aim is to gain better physical understanding of the AGN phenomenon as is emerging from high resolution and S/N observations and understand the dynamics of highly ionized gas in those systems. In addition, we wish to quantify the mass loss rate from these objects which have been claimed to have a paramounting effect on the environment of such objects. 

In this paper we consider a sample of X-ray bright,
nearby type-I AGN (Seyfert 1-1.5 galaxies) for which high quality
grating spectra are available and the mass of the central
black-hole has been measured. We improve upon the model of CN05
and apply it to study HIG outflows in those objects. The paper is
organized as follows: In \S 2 we present the sample of objects
used in this work and discuss their properties. Section 3 outlines
the basics of our model and the improvements and generalizations
made to the CN05 scheme. Results of our spectral modelling
pertaining to individual objects are presented in \S 4. We discuss
the implications of our results for the unification scheme of AGN
and AGN interaction with their environment in \S 5. Summary
follows in \S 6.

\section{The sample}

Our sample was constructed having in mind the Netzer et al. (2003)
and CN05 spectral analysis method. This method uses the silicon
and sulphur lines near 2\,keV to constrain the properties of the
ionized outflow over a wide range of ionization levels. This
enables one to eliminate many of the uncertainties and degeneracies
associated with the intrinsic continuum shape at soft X-rays (e.g.,
Netzer et al. 2003). Also, the CN05 results suggest that, at least
for NGC\,3783, most of the mass loss is due to an extremely
ionized component of the flow. Such gas is, however, much harder
to detect and requires good signal to noise (S/N) and high
resolution observations in the hard X-ray band. Currently, these
wavelengths are best covered by the {\it Chandra} gratings (the
{\it XMM/RGS} has negligible effective area shortward of $\sim
12\AA$). For this reason our sample is essentially flux limited and
is restricted to relatively nearby objects that have been observed
by the {\it Chandra} high energy transmission grating (HETG); we
note that despite its large effective area, the LETG is not
favored for this study due to its very low resolution at short
wavelengths (see also \S 4.1). We have thus searched the {\it Chandra} archive for
bright AGN with long grating exposure times. Our search yielded
five objects (including NGC\,3783 which is included here only for
completeness purposes and is discussed in CN05) that are listed in
table 1. For each object in our sample we have collected
all archived {\it Chandra/HETG} (and in some cases also LETG) observations and reduced those
using  CIAO (v3.1) and CALDB (v3.1), using the standard CXC threads. In some case we also
analyzed {\it XMM} data to gain additional information concerning
the source although these spectra have not been used to derive the outflow properties.
({\it XMM} data were reduced using the Science Analysis Software,
SAS v.5.3.0 with the standard processing chains. For the detailed spectral analysis (see section 4) we have included only HETG data and, where possible, combined several exposures to increase the S/N, after verifying they have a similar spectrum (see table 1). 

Some relevant physical properties for each object in our sample
are given in table 1. An important parameter for any
dynamical modelling and, in particular, to the CN05 model, is the
mass of the central black hole, $M_{\rm BH}$, assumed to dominate
in the inner regions of the AGN (see however Sh\"{o}del et al.
2003 and Everett \& Murray 2007). Thus, it is important that the black hole mass be known to
a good precision. Currently, the best way this is done for AGN is
by reverberation mapping techniques (e.g., Kaspi et al. 2000). For
all objects in our sample but one, reverberation masses were taken
from Peterson et al. (2004). For one object in our sample
(MCG-6-30-15) the mass was estimated from variability considerations and from the
mass-velocity dispersion relation (McHardy et al. 2005).

\begin{figure}
\plotone{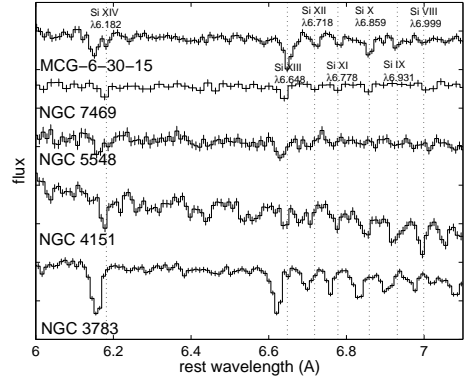} \caption{The X-ray band containing the inner
shell silicon line spectrum for all the objects in our sample.
Clear absorption signatures due to highly ionized gas
(specifically due to \ion{Si}{14}\,$\lambda 6.187$) are evident in
all objects regardless of their classification and/or prior
evidence for the existence of ``warm absorbing'' gas (George et
al. 1998). Also, the blueshift of the lines with respect to the systemic
velocity of the system indicates gas which is outflowing. We note that the
spectrum of NGC\,7469 is the combined spectrum from two
observations (after correcting for the slight difference in flux
level) and rebinned to to increase the S/N.} 
\label{fig1}
\end{figure}

\begin{figure}
\plotone{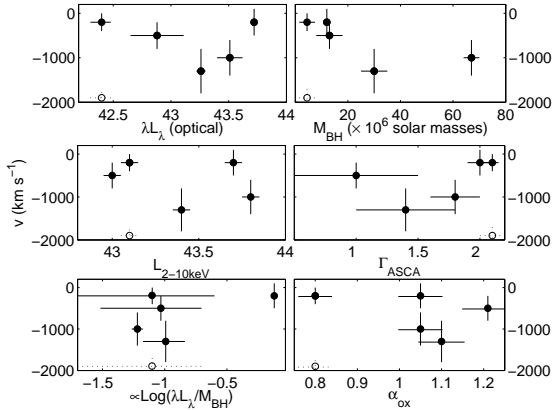} \caption{The
outflow velocity as inferred from the \ion{Si}{14}\,$\lambda
6.128$ absorption line for all objects in our sample as a function
of object's properties. Given the few number of points and the
measurement uncertainties, no significant correlations can be seen
in our sample. Nevertheless, it is clear that there is a wide
range of outflow velocities for a given Eddington ratio.  
Uncertainties in the measured line velocity
include wavelength position uncertainties as well as the effect of
non-Gaussian line profiles (see text). It should be noted that
redshifts toward individual may be uncertain by about $100~{\rm
km~s^{-1}}$ (see Crenshaw et al. 1999).} \label{fig2}
\end{figure}

\begin{table*}
\begin{center}
{\sc TABLE 1 \\ The Sample}
\vskip 4pt
\begin{tabular}{llllllll}
\tableline
\tableline
Object & z & $M_{\rm BH}^\star$ & $\lambda L _\lambda^\star$ & $L_x$ & $\Gamma_{\rm ASCA}$ & $\alpha_{ox}$ & HETG time\\
 & & ${\rm 10^6\,M_\odot}$ & ${\rm erg~s^{-1}}$ & ${\rm erg~s^{-1}}$ & & & ksec \\
\tableline
NGC\,3783 & 0.00973 & $30\pm5$ & $43.26$ & $43.4$ & $1.8$ & $1.1$ & 680 (ObsID 2090, 2091, 2092, \& 2094)\\
NGC\,5548 & 0.01717 & $67\pm3$ & $43.51$ & $43.8$ & $1.9$ & $1.05$ & 150 (ObsID 3046)\\
NGC\,7469 & 0.01632 & $12\pm1$ & $43.72$ & $43.7$ & $2.02$ & $1.05$ & 150 (ObsID 2956 \& 3147)\\
NGC\,4151 & 0.00332 & $13\pm5$ & $42.88$ & $43.0$ & $1.39$ & $1.2$ & 250 (ObsID 3052 \& 3480)\\
MCG-6-30-15 & 0.00775 & $4.5\pm3^\dagger$ & $42.4$ & $43.1$ & $2.1$ & $0.8$ & 500 (ObsID 4759, 4760, 4761, \& 4762)\\
\tableline
Total HETG time & & & & & & & $\sim 1700$ \\
\tableline
\end{tabular}
\vskip 2pt
\parbox{5.2in}{ 
\small\baselineskip 9pt
\footnotesize
\indent
$^\star$ Taken from Peterson et al. (2004); Redshifts taken from Peterson et al. these differ from other values given in the literature by no more than 120 km/s \newline
$^\dagger$ Estimated from variability and the black-hole--bulge relation in galaxies; see McHardy et al. 2005
}
\end{center}
\end{table*}

Figure \ref{fig1} shows the silicon band in the HETG spectra for all
objects in our sample. Line signatures
from highly ionized gas (specifically that of the \ion{Si}{14}
line near $\lambda 6.128$) are easily detectable. Thus,
highly ionized absorbing gas is evidently present in {\it all}
objects. This is an important finding since the sample was not pre-selected
to detect such gas but rather to observe gas at much lower
ionization levels pertaining to the so called ``warm absorbing''
phenomenon (e.g., George et al. 1998) and colder, more neutral, soft X-ray absorber. 
Gas at such high ionization
levels would have evaded detection by previous observatories such
as {\it ASCA}, but is clearly seen here. Moreover, the presence of
this extremely ionized gas component in all objects reveals the
existence of an important constituent of AGN. Indeed, the presence of such
highly ionized gas in cases where outflowing warm absorbers are
detected is naturally explained by the thermal wind picture of CN05. It is nevertheless
unclear whether
gas with even higher ionization levels is also outflowing in those
objects (see e.g., Reeves et al. 2004) as the effective area and
resolution of {\it Chandra} gratings is insufficient at the high
energy band and the analysis of CCD spectra is of much lower resolution and is plagued with uncertainties related to photon redistribution effects.

The advantage of X-ray grating over non-grating spectra is that it
allows us to unambiguously identify spectral features and measure
absorption line velocities to great accuracy. We have measured the
velocities of various silicon lines for all the objects in our
sample and looked for correlations with other AGN properties;
these are shown in figure \ref{fig2}. Clearly, no firm conclusions
can be drawn from the small sample at hand. Nevertheless, it is
indicative that the outflow velocity (as inferred from the
\ion{Si}{14}\,$\lambda 6.128$) is not intimately related to the
Eddington ratio of the source and that AGN with apparently similar
Eddington ratios can have HIG traveling with a wide range of
velocities. The data hints at a possible trend between
the flow velocity and the black hole mass. We caution however that
this is dubious given the uncertainties involved and the
small number statistics (note also that the appearance of a high
velocity component in MCG-6-30-15 which does not agree with this trend
and is further discussed in \S 4.4). No
clear trends are seen as a function of the optical to X-ray
luminosity slope, $\alpha_{ox}$, the X-ray luminosity, the X-ray
photon index, $\Gamma_{\rm ASCA}$ (George et al. 1998), or the
optical luminosity, $\lambda L_\lambda$ (taken from Peterson et
al. 2004 and for MCG-6-30-15 from George et al. 1998 using a
typical type-I UV to optical conversion factor). Similar results
are obtained for the outflow velocities inferred from other,
non-blended, lines in the {\it Chandra/HETG} bandpass. In short,
no  clue as to the physical mechanism driving such flows in AGN
may be drawn from the small sample at hand. The dependence of the
other flow properties such as the column density, total opacity,
ionization level, and temperature, is very model dependent and is
not discussed here (see Blustin et al. 2005).

\section{The Model}

Here we outline the model used in this work to derive the physical
properties of the outflowing gas. This is based on a model
extensively described in CN05 and applied to the outlflowing gas
in NGC\,3783. Thus, we mention here only briefly the basic model
ingredients and refer the reader to CN05 for
an elaborate discussion and justification of the formalism.

The model used here is that for a multiphase, radiation and
thermal pressure driven flow. The flow spans a range of
densities at every location $r$ (perhaps due to thermal
instabilities or the onset of turbulence; CN05) and is assumed to obey a
density-scale distribution of the form
\begin{equation}
\rho \propto \xi^\beta
\label{density_spectrum}
\end{equation}
where $\beta$ is the powerlaw index (assumed constant throughout
the flow and its value is to be determined from observations) and
$\xi$ is a measure of length. The ionization and thermal properties of each phase in the flow are self-consistently calculated using the photoionization code {\sc ion} (e.g., Netzer 1996) which includes all relevant heating and cooling mechanisms. Self-shielding is self-consistently accounted for by our calculation scheme and we use the escape probability method to calculate emission line transfer either in static or differentially expanding media (see below). Once the ionization structure of the absorbing gas is obtained, the radiation pressure acceleration, $a_{\rm rad}$ is calculated taking into account all absorption and scattering processes such as bound-bound, bound-free, free-free absorption, as well as Compton scattering (see Chelouche \& Netzer 2003). Complete knowledge of the gas thermal and ionization properties allows us to go one step further beyond pure photoionization models and consider the kinematic part of the problem at hand.

\begin{figure}
\plotone{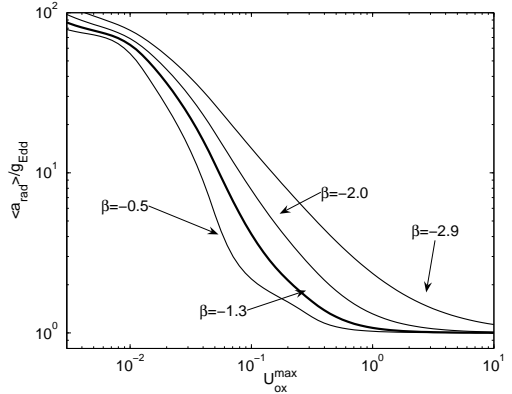} \caption{The average radiation
pressure acceleration relative to gravity (for a source emitting at its
Eddington rate) as a function of the maximum ionization parameter
in the flow, $U_{\rm ox}^{\rm max}$ and for several values of the
density spectrum powerlaw index $\beta$. The minimum ionization
paramter, $U_{\rm ox}^{\rm min}$ in the flow was arbitrarily set
to $10^{-4}$ and its exact value has little effect on the
results. Clearly, the mean radiation pressure acceleration is not much
larger than the Compton pressure for plausible values of the
parameters (see e.g., CN05, and table 2). This implies that radiation
pressure force on the HIG kinematics may be neglected in objects emitting 
below their Eddington rate.} \label{fig3}
\end{figure}

We first define the continuity condition for the outflowing gas. As shown in CN05, for a flow with a fixed density contrast ratio at all locations, the
continuity condition for the most dilute component, i.e., that
with the highest ionization parameter; $U_{\rm ox}^{\rm max}$ and lowest mass density, $\rho_{\rm min}(r)$, is similar to the
spherical expansion case and
\begin{equation}
\rho_{\rm min}(r) \propto r^{-2}v(r)^{-1}.
\label{cont}
\end{equation}
The flow velocity, $v(r)$, is obtained by solving the general equation of
motion for the multi-phase wind (assuming all phases are
dynamically  coupled) which takes the form,
\begin{equation}
\frac{1}{v}\frac{dv}{dr}=\frac{1}{v^2-(1+\beta/3)\gamma v_s^2}\left [
  \frac{2(1+\beta/3)\gamma v_s^2}{r}-\frac{GM_{\rm BH}^{\rm eff}}{r^2}
  \right ],
\label{eqnmot}
\end{equation}
where $v_s$ the sound speed at the critical point, $r_c$ (see
below) and $\gamma$ the polytropic index (i.e., the gas temperature, $T\propto \rho^{\gamma-1}$). We note that $\gamma$ is
merely a parametrization which enables a relatively
straightforward integration of the equation of motion and whose
value depends, as far as the dynamical problem is concerned, on
the heating and cooling processes of the most ionized flow phase;
e.g., photo-heating, radiative cooling, adiabatic expansion, and
possibly other heating mechanisms such as dissipation of turbulent
energy (e.g., CN05). $\gamma = 1$ corresponds to isothermal
gas. We define the effective black hole mass, $M_{\rm BH}^{\rm
eff}$ as
\begin{equation}
M_{\rm BH}^{\rm eff}\equiv M_{\rm BH}\left ( 1- \frac{\left < a_{\rm rad} \right > }{g} \right )
\label{mbheff}
\end{equation}
where $g$ is the gravity (for a non- or a highly sub-Keplerian
rotating flow, $M_{\rm BH}^{\rm eff}$ has only a very weak
dependence on $r$; CN05), and $\left < a_{\rm rad} \right >$ is
the average radiation pressure acceleration at a given location in the flow defined
as
\begin{equation}
\left < a_{\rm rad}(r) \right > = \frac{\int dm a_{\rm rad}(\rho;r)}{\int dm },
\label{arad}
\end{equation}
where $dm=\rho dV(\rho)$ is the contribution to the mass from each
flow phase.  As $\left < a_{\rm rad} \right >$ depends on the local
properties of the flow such as the ionization, temperature,
density contrast, it is not known a-priori and must be solved
 in conjunction with the equation of motion.
Figure \ref{fig3} shows $\left < a_{\rm rad}\right >/g$ for a
source emitting at its Eddington rate ($g=g_{\rm Edd}$). Clearly,
for all plausible values of $\beta$ and $U_{\rm ox}^{\rm max}$,
the radiation pressure acceleration barely exceeds that due to
Compton radiation pressure (i.e., unity in the adopted units).
This property allows this term to be neglected in the CN05 analysis of
NGC\,3783 due to the low Eddington rate of the source ($g\gg
g_{\rm Edd}$ hence $\left < a_{\rm rad} \right
>/g \ll 1$). Nevertheless, the radiation pressure acceleration can have a
non-negligible effect on the dynamics for sources emitting close
to their Eddington rate (such may be the case of NGC\,7469 in our
sample).

The equation of motion requires that a steady-state solution be
initially subsonic and reach super-sonic velocities while passing
through a critical point where  both the numerator and the
denominator vanish (e.g., Parker 1965).  This point is given by
\begin{equation}
r_c=\left [ \left ( 1+\beta/3 \right ) \gamma \right ]^{-1}
\frac{GM_{\rm BH}^{\rm eff}}{2v_s^2},~~ v(r_c)=v_s\sqrt{\left ( 1+
\beta/3 \right)\gamma}. \label{crit}
\end{equation}

Once the flow density and velocity are determined by the model at
some location along flow lines and assuming a divergence free flow
(i.e., there are no mass sinks or sources) the mass loss rate is
given by (CN05),
\begin{equation}
\dot{M}=C_{\rm global}\frac{8\pi r^2\rho_{\rm min} v}{\beta+2}\frac{1-(\rho_{\rm max}/\rho_{\rm min})^{(\beta+2)/\beta}}{1-(\rho_{\rm max}/\rho_{\rm min})^{2/\beta}}.
\label{dotm1}
\end{equation}
where $C_{\rm global}$ is the global (in units of $4\pi$) covering factor.

The equation of motion (eq. \ref{eqnmot}) combined with the
continuity condition (eq. \ref{cont}) must be solved in
conjunction with the photoionization and thermal equations to
yield a self-consistent solution for the structure and dynamics of
the outflowing gas. We note that self-shielding can be important
and is accounted for by our model. The model also includes the
effects of adiabatic cooling as well as a possible additional
heating source in the form of dissipation of turbulent energy. We
refer the reader to CN05 for the detailed layout of the model and
our calculation scheme. Throughout this work we assume solar 
composition gas (see Arav et al. 2001, 2007, Netzer et al. 2003 and references 
therein).
Once the velocity profile and the flow ionization and thermal
structure are known for every phase at every location, we
calculate the transmitted spectrum through the flow including all
important lines and edges while accounting for the differential
expansion of the flow via the CN05 formalism. The synthetic spectrum is then convolved
with the instrumental kernel (assumed Gaussian) and is compared to
the data. 

A comparison of the model prediction to the observed line profiles as well as to
the continuum shape allows us to constrain the parameters of the
model which include the location at which the flow crosses our
line of sight, $r_0$ (this needs not be equal to $r_c$ as lines can
be detached; see e.g., Arav 1996 and CN05), the range of
ionization parameters occupied by the flow at $r_0$, $[U_{\rm
ox}^{\rm min}, U_{\rm ox}^{\rm max}]$, and $\beta$. Once these
parameters are set  the entire solution is determined,
hence the absorption spectrum. We note that in addition to the above parameters 
that characterize the outflow,  the object's luminosity, 
black hole mass, and its intrinsic spectral energy distribution
must also be specified. 

Our fitting procedure is fully interactive due to the complexity
of the model. We first look for the highest ionization species
identifiable in the absorption spectrum. These allow us to estimate or put a
lower bound on $U_{\rm ox}^{\rm max}$. This is important since the
highest ionization component sets the dynamics of the entire
multi-phase flow (CN05). From the low ionization silicon lines (and to
some extent also the continuum shape at soft X-rays; but see \S
4.3 , 4.4), we can estimate $U_{\rm ox}^{\rm min}$  and set the
density contrast ratio in the flow. Estimating $U_{\rm ox}^{\rm min}$
from observations is more challenging than deducing the value of
$U_{\rm ox}^{\rm max}$ due to uncertainties in the shape of the soft X-ray 
continuum, yet this has a negligible effect on the derived mass loss rates 
and the kinetics luminosities. The powerlaw index $\beta$ can be constrained 
by its combined effect on the spectral shape and on the gas dynamics (CN05).

Once an acceptable fit is found (given the criteria used in CN05), the physical 
properties of the outflowing gas and the
uncertainties on individual parameters are obtained by manually changing their values and
then searching for an adequate agreement between the model and the data by varying all
other parameters of the model (e.g., $\beta,~r_c,~r_0, C_{\rm
global}$, $U_{\rm ox}^{\rm max}$, and the density contrast). Thus, the quoted errors are not
independent among different parameters and inter-correlations
exist between them (see CN05 for further details).

\subsection{Emission line treatment}

Once an adequate model is found based on a global agreement with
the absorption data and the flow structure and dynamics are
constrained, we calculate the emission spectrum from the
outflowing gas. For this purpose we assume that the global flow
structure for all lines of sight is similar to the one which is
intercepted by us. While this may not be true in general (e.g.,
Kaspi et al. 2004), it is the only viable approach to the problem at this 
stage. Calculating the emission spectrum from a differentially expanding
medium is an extremely complicated task and requires detailed
ray-tracing calculations for hundreds of lines. Here we take a
very simplified approach which, despite its limitations (see
below), accounts in a better way for the emission lines from such
flows.

\begin{figure}
\plotone{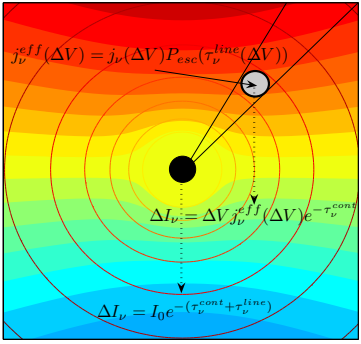} \caption{Schematics of the
iso-velocity contours of a wind model as seen by an observer at
the bottom of the diagram. Also shown is a parcel of the
outflowing gas which emits line photons. Within the parcel we
assume the effective line emissivity, $j_\nu^{\rm eff}$ to be the
product of the intrinsic emissivity, $j_\nu$ and the static escape
probability, $P_{\rm esc}$ (assuming the static case applies; see
text). Continuum attenuation is then calculated along the photon
path from its creation point to the observer.} \label{fig4}
\end{figure}

\begin{figure}
\plotone{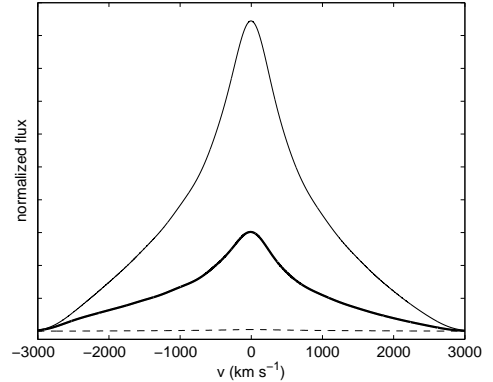} \caption{The emission line
profiles from a spherically symmetric expanding flow when
calculated using the full scheme described in text (thin line)
and compared to the case in which no continuum absorption of line
photons is assumed (thick line).  Clearly, the difference between 
the two methods is considerable and detailed calculations are necessary
to constrain the global covering factor of the flow. We also
note the asymmetry of the fully calculated line profile due to
stronger absorption of the receding part of the flow.}
\label{fig5}
\end{figure}

Previous works (e.g., Netzer et al. 2003) have calculated the
emission spectrum assuming stationary gas. Nevertheless, photon
scattering in differentially expanding medium is limited to
sections of the flow having similar velocities up to roughly one
"thermal" width (by "thermal" we mean either purely thermal or
that which is set by small scale turbulent velocity field, so
called micro-turbulence). For larger velocity differences, the
absorption cross-section is considerably reduced and photon
scattering is less important. This is the basic notion behind the
well established Sobolev approximation which we use here for
treating emission lines. We therefore divide the flow into
segments in 3D space such that they are independent of one another
with respect to photon scattering in lines. We assume that different transitions do not
overlap in energy space which is good enough an approximation for the relatively
low velocities considered here (e.g., Chelouche \& Netzer 2003;
see however the \ion{O}{6}$\lambda21.87$ line problem, Netzer et
al. 2003). Naturally, different lines have different thermal
widths, and this is further complicated by the possible emergence
of turbulence (e.g., CN05). We take the simplified approach in
which all lines are affected similarly and use the gas sound speed of
the highest ionization component as a measure of the line width
for all lines. Within each flow segment we calculate the emergent
spectrum assuming it to be stationary in its own frame (i.e., we
neglect differential expansion within individual segments). We 
use the static approach (e.g., Netzer 1996 and references therein) 
in the sub-sonic part of the wind. While this treatment of the emission line
transfer is approximate even within the framework of the escape
probability method, the
uncertainties associated with the unknown global geometry of the
outflowing gas are likely to dominate (e.g., Bonilha et al. 1979). The
emission spectrum is calculated by summing over the contribution
from all flow segments and accounting for the velocity shifts
between them. A major concern using this scheme is the effect of
continuum absorption on the line flux (see figure \ref{fig4}).
Contrary to lines, this process is non-resonant; i.e., photons do
not scatter but are absorbed by gas lying in our line-of-sight to
the point of creation of the photon. While for most cases in our
sample this effect causes line fluxes to differ by roughly 20\%
(this was verified by calculating the emission spectra assuming
complete absorption by the flow and no absorption at all), for one
object in our sample (NGC\,4151) the effect turns out to be
substantial.

The wavelength dependent continuum optical depth for a photon
created at some location $r',~\theta'$ within the flow is
\begin{equation}
\begin{array}{l}
\displaystyle \tau^{\rm cont}_\nu(r',\theta')=\left [ -{\rm sign}({\rm cos} \theta')\int_{r'}^{r'{\rm sin}\theta'}+\int_{r'{\rm sin}\theta'}^\infty \right ] \\
\displaystyle  ~~~~~~~~~~~~~~~~~~~~~~~~\times \left (
\frac{d\tau_\nu^{\rm cont}}{dr} \right ) \sqrt{1-\frac{r'^2}{r^2}{\rm
sin}^2\theta'}dr
\end{array}
\label{tau}
\end{equation}
Once the flow opacity profile is constrained by our model for a single line of sight, and assuming spherical symmetry, the
above integral can be readily calculated numerically at every
frequency, and the process can be repeated for all emission lines.
Depending on the number of zones required (i.e., on the
sound-speed of the gas), the calculation may take several hours on
a Linux workstation (we note that, for the specific case in which $d\tau/dr$ is a
powerlaw of $r$, a semi-analytic form for the integral exists
employing the Hypergeometric functions). Figure \ref{fig5}
demonstrates the different emission line fluxes and profiles
obtained by assuming no/full continuum absorption of the emitting
region, and accounting for the flow geometry in full as described
above. The line intensity is different as well as the line
profile being slightly asymmetric in the full calculation due to
higher flow opacity toward the receding side compared to the
approaching side. The full radiative transfer yields emission
lines with intermediate intensities due to geometry affecting the
extended emission region. 

When calculating the integrated emission spectrum we include the
contribution of all phases of the flow at every location
and the emission/absorption spectra
are combined. We note that due to the unknown geometry of the
emission region with respect to our viewing angle, spherical
symmetry is assumed. By changing the global covering factor of the emitting gas, 
$C_{\rm global}$, we do not alter the geometry of the flow but rather rescale its 
overall contribution to the flux.
Again, this approach is probably a crude over-simplification yet we 
believe it is the only viable assumption at this stage.  The resulting emission/absorption 
spectrum depends on the parameters of the model and a comparison with the spectral
data may be used to deduce their values.

\section{Results}

In this section we discuss our results pertaining to the spectra of individual objects in our sample.  The detailed
analysis for each object is presented separately since, despite
their overall similarities, their properties and their spectra can
differ substantially. The implications of our results are discussed
in section 5. 

\subsection{NGC\,5548}

\begin{figure}
\plotone{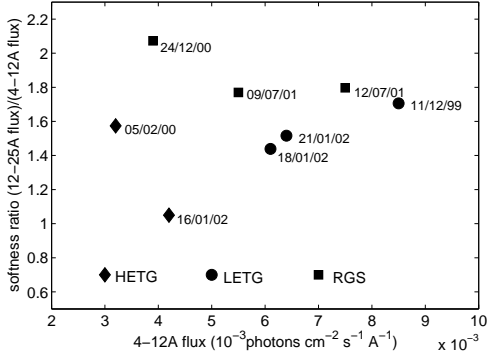} \caption{The softness ratio as a
function of the hard X-ray flux for eight X-ray grating
observations of NGC\,5548 (three {\it Chandra/LETG}, two {\it
Chandra/HETG}, and two {\it XMM/RGS}). Clearly, there is no
correlation between the softness ratio and the flux level. This is
similar to the behavior of NGC\,3783 which was reported by Netzer
et al. (2003).} \label{fig6}
\end{figure}

NGC\,5548 is a typical type-I Seyfert galaxy and has perhaps the
best studied warm absorbing gas after NGC\,3783 (Kaastra et al.
2000). The hard X-ray ($1-10$\,keV) photon slope is known to vary
in the range 1.5-1.7 (Chiang \& Blaes 2003) with a mean value of
1.6 (Kaastra et al. 2004). There is evidence for a steep soft
X-ray slope (Kaastra et al. 2004) reminiscent of the case of NGC\,3783 (Netzer
et al. 2003). The optical to X-ray spectral slope, $\alpha_{ox}$ is also
similar to that of NGC\,3783 (see table 1). Thus, there is no
direct evidence that the ionizing spectra of NGC\,5548 and
NGC\,3783 differ substantially and, for the purpose of our model, we 
assume they are identical and, therefore, use the ionizing continuum shape
defined in Netzer et al. (2003). We find the bolometric luminosity
of the NGC\,5548 (given its mean 2-10\,keV X-ray flux reported by
Steenbrugge et al. 2003) to be $\sim 7\times 10^{44}~{\rm
erg~s^{-1}}$ and conclude that the Eddington ratio is $\sim
0.1$, i.e., comparable to that of NGC\,3783.

\begin{figure}
\plotone{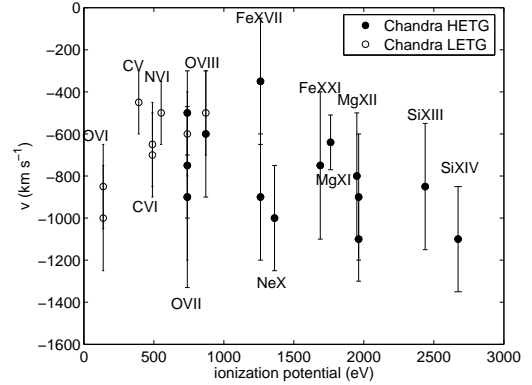}
\caption{Line velocity as measured from {\it Chandra} HETG and
LETG observations of NGC\,5548. Our measurements do not show a significant correlation between 
the velocity of the line transition(s) and the ionization potential of the ion giving rise to those transitions. This is similar to the case of NGC\,3783 (see however Steenbrugge et al. 2005 and text).} \label{fig7}
\end{figure}

\begin{figure}
\plotone{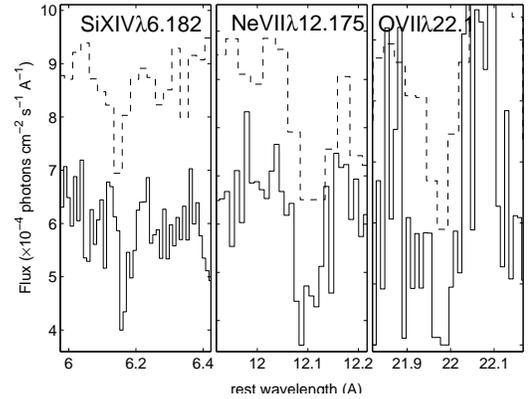}
\caption{A comparison between the HETG (solid line) and LETG (dashed line) data for NGC\,5548 which were taken at different epochs. There is an evident shift of high ionization lines to higher velocities compared to low ionization lines in the LETG data. This may be either due to changes in the flow opacity at high ionization levels during the LETG epoch or due to wavelength calibration issues between the instruments (see also Steenbrugge et al. 2005.} \label{fig7.5}
\end{figure}

\begin{figure*}
\plotone{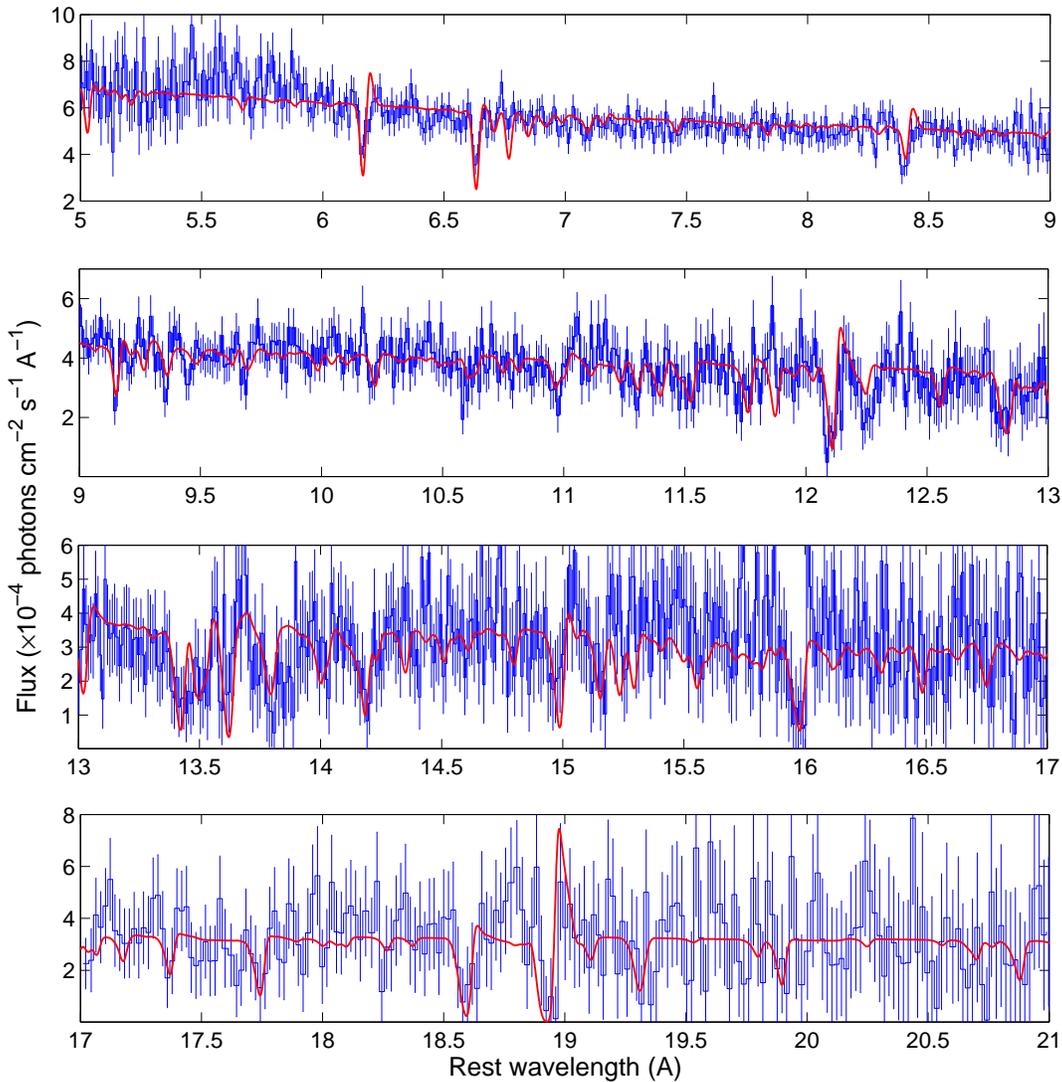}
\caption{A spectral model for a thermally driven wind overlaid on  the {\it Chandra/HETG} spectra
of NGC\,5548. Clearly, the model reproduces most absorption line features and traces well 
the continuum shape over almost a decade in photon frequency.  Some residuals 
near 15\AA are possibly due to uncertainties in the  di-electronic recombination
coefficients of iron (e.g., Netzer 2004). Contribution of the outflow to the emission is clearly evident. Note that data binning is different between the panels.}
\label{fig8}
\end{figure*}

NGC\,5548  has been observed three times by {\it XMM} (e.g.,
Steenbrugge et al. 2003, Pounds et al. 2003) and five times by
{\it Chandra} (two {\it HETG} and three {\it LETG} observations;
e.g., Kaastra et al. 2002). The object was found to vary by nearly
a factor four in luminosity over a period of three years. The
spectral shape (expressed here by the softness ratio; see figure
\ref{fig6}) also varied by $\sim 40\%$ during that period with no
obvious correlation with the flux level (see also Kaastra et al.
2004). This behaviour is very similar to the one reported by
Netzer et al. (2003) for NGC\,3783 and suggests a common physical
mechanism that drives these phenomena in both objects. From
spectral analysis of X-ray absorption features, Steenbrugge et al.
(2003, 2005) conclude that the HIG is highly stratified and spans
a large range of temperatures ($\sim 10^4-10^6$\,K). A similar
conclusion was reached by Netzer et al. (2003) and CN05 for
the X-ray outflow in NGC\,3783. Detailed photoionization modelling
of the HIG in both objects suggest that low ionization gas
components have smaller column densities than the high ionization
ones (with a qualitatively similar scale dependence obtained for
both objects; c.f. Steenbrugge et al. 2005 and CN05). The outflow
velocity in both objects is also similar as has been verified by
observations of several absorption lines (cf. Kaspi et al. 2002
and Steenbrugge et al. 2005). Neither the HIG in NGC\,5548, nor
that in NGC\,3783 show clear response with respect to variations
of the continuum flux level of the source implying, by recombination
time-scale arguments, that the flow is located on parsec scales.

We have measured the line velocities for the high ionization
species from the {\it Chandra} HETG and LETG observations (see
figure \ref{fig7}). For high ionization lines we rely on measurements from 
the HETG which has a higher spectral resolution.  We find no clear 
trend  between the ionization level (expressed here using the photoionization
threshold energy of the ion) and the velocity.  We compared the LETG and HETG spectrum (see figure \ref{fig7.5}) and find that high ionization LETG lines are more blueshifted compared to the HETG lines. The cause for this may be either due to changes in the flow opacity with time which are manifested only in relatively high ionization lines (e.g., \ion{Si}{14}$\lambda 6.182$ and \ion{Si}{11}$\lambda 6.778$), or calibration issues. While the underlying reason for the discrepancy is unclear, this is probably the cause for the different result in our work compared to Steenbrugge et al. (2005).  

Motivated by the above similarities between NGC\,5548 and
NGC\,3783 we attempted to explain the highly ionized outflow in
NGC\,5548 with a model similar to the one presented in CN05. Our
fitting procedure follows the scheme presented in \S 3 and discussed in CN05.  Here too,
we find the effect of radiation pressure acceleration on the flow
dynamics to be negligible (i.e., $\left < a_{\rm rad}\right >/g
\ll 1$; see figure \ref{fig3}). The best-fit underlying photon
powerlaw index is $1.6$ (i.e., similar to that of NGC\,3783) and
our scheme converges to an acceptable spectral model shown in figure
\ref{fig8} and whose parameters are given in table 2. We have
tried models with and without additional heating sources, like in
the case of NGC\,3783. We arrive to the same conclusions as in
CN05 regarding the necessity for an additional heating source,
perhaps due to the dissipation of turbulence to counter-balance
the effect of adiabatic cooling. Excluding an additional input of
heat will result in a similar overall fit to the continuum shape only 
with narrower and less blueshifted lines which are less favored
by the data (see also CN05).

Clearly, the model traces well the continuum and lines
of the more abundant elements but over-estimates the absorption
due to \ion{Si}{13} and \ion{Si}{14} at around 6.5\AA. There are
several possible reasons for that: the first is that the abundance
of silicon is somewhat lower than solar. Our calculations show
that the data are consistent with the abundance of silicon being
roughly 30\% solar. Evidence for deviations from solar composition
have been previously suggested to explain the UV absorption line
spectrum of quasar outflows (e.g., Arav et al. 2001, 2007). Troughs may
appear deeper in our model due to under-estimated emission whose
normalization is derived from global fitting.  It is also possible
that the over-estimation of the absorption by highly ionized
silicon lines is related to time-dependent ionization. For the
densities implied by our model ($\sim 10^3-10^6~{\rm cm^{-3}}$ at
$r_0$), we find that the equilibration timescales are of order a
week (Krolik \& Kriss 2000; correcting for the different density and
flux). Hence, the ionization level depends on
the long-term history of the flux and SED of the source.
Unfortunately, we do not have such data at hand. Nevertheless,
there is a clear indication that the object brightened in the hard
band by about 50\% over a period of 2 days (see figure \ref{fig6})
and is expected to result in some deviation from a pure ionization
equilibrium case. These would be again small if the
equilibration timescale is much longer than the variation
timescale; indeed comparing the LETG observation taken two days
prior to the first HETG observation shows no significant changes in
the lines apart from what we suspect to be due to calibration effects. 
Last is the possibility of thermal instability driving
some of the gas to higher temperatures so that more of the gas
becomes transparent and not contribute to the silicon line
opacity. The time-dependence of thermal instability phenomenon has
not been studied in detail for such gas and its proper treatment
is beyond the scope of this paper.

In addition to the above discrepancy, our model is slightly below
the data around $15.5\AA$. This problem was already noted by CN05
and likely results from uncertainties in the di-electronic recombination
rates of iron (e.g., Netzer 2003) as well as uncertainties in the
properties of the unresolved transition array (UTA) of iron whose
most up-to-date values (Gu et al. 2006) have not been incorporated
into our calculation scheme.

The calculation of the mass loss rate requires knowledge of the
global covering fraction of the absorber, $C_{\rm global}$. This
can be obtained by fitting for the emission lines. This is
accomplished here by assuming that the flux level of the source
represents  its long-term average and that it has not varied
significantly over the equilibration timescale of the ionized gas
and the light crossing time of the emission region (as discussed
above this assumption may not be well justified yet it is the only
viable one at this stage). Our model suggests that the data are
consistent with the flow fully covering the ionizing source and is
therefore different from the case of NGC\,3783 (Netzer et al.
2003). This is expected given the similar equivalent width of the
emission lines for the two objects  (e.g., oxygen lines, compare
Kaspi et al. 2003 and Steenbrugge et al. 2005) while noting that the column
density of the absorber in NGC\,5548 is smaller by a factor of a
few.  The obtained mass loss rate and the kinetic
luminosity associated with the flow are given in table 2.

\begin{table*}
\begin{center}
{\sc TABLE 2 \\ Best-fit model parameters}
\vskip 4pt
\begin{tabular}{lrrrrrr}
\tableline
\tableline
Parameter & NGC\,3783 & NGC\,5548 & NGC\,7469$^{(a)}$ & NGC\,4151 & MCG-6-30-15 (low/high) \\
\tableline
Critical distance $r_c$ [$10^{18}$\,cm] & $2.0\pm 1.5$ & $7\pm2$ & $1.0\pm 0.5$ & $1.5\pm0.8$ & $(8\pm 8/0.2\pm0.15)$\\
Line-of-sight crossing distance $r_0$ [$10^{18}$\,cm] & $5\pm2$ & $6\pm 2$ & $1.0\pm0.5$ & $0.3\pm0.2$ & $(10\pm 7/2\pm 1)$\\
Maximum ionization parameter at $r_0$ [${\rm log}(U_{\rm ox}^{\rm max})$] & $0.5\pm0.3$ & $0.1\pm0.3$ & $0.7\pm0.7$ & $-0.3\pm0.5$ & $(-0.2\pm 0.5/0.5\pm 0.3)$\\
Minimum ionization parameter at $r_0$ [${\rm log}(U_{\rm ox}^{\rm min})$] & $-3.7\pm1.5$ & $-4\pm1$ & $-3.0\pm1.5$ & $-4\pm1$ & $(-2.5\pm 1.0/-3\pm 1)$\\
Density spectrum index  ($\beta$)  & $-1.2\pm0.5$ &  $-1.4\pm 0.6$& $-1.3\pm0.6$ & $-1.4\pm0.3$ & $(-1.3\pm0.6/-1.3\pm 0.5)$\\
Global covering factor$^{(b)}$, $C_{\rm global}$ & $0.2$ & $1$ & $1$ & $1$ & $(0.7/1.0)$\\
\hline Mass loss rate [${\rm log(M_{\odot}~yr^{-1})}$], &
$-1.5\pm0.5 $& $-0.2\pm 0.6$ & $-2.0\pm 0.8$ & $-3.2\pm 0.7$ & $(-1.0\pm0.8/-2.2\pm0.7)$\\
Mass accretion rate $(=90\lambda L_\lambda/c^2)~[{\rm
log(M_{\odot}~yr^{-1})}$] & $-1.7$ & $-1.5$ & $-1.3$ & $-2.2$ &  $-2.6$ \\
Relative Kinetic Luminosity $(L_{\rm kin}/L)^{(c)}$ [per cent] & $0.01$ & $0.1$ & $0.0001$ & $0.0001$ & $0.02$ \\
\tableline
\end{tabular}
\vskip 2pt
\parbox{6.0in}{ 
\small\baselineskip 9pt \footnotesize \indent 
$^{(a)}$ The effective mass of NGC\,7469 has a large uncertainty. The quoted errors assume the object emits at about 90\% of its Eddington rate. While the object could, in principal, emit more efficiently, we find it less likely for reasons described in text \\
$^{(b)}$ We note
that the global covering fraction was derived assuming spherical
symmetry for the absorbing gas and assuming that the outflowing
matter is solely responsible to the emission. If other components
are responsible for the emission then $C_{\rm global}$ reported
here is only an upper limit. The covering factor for NGC\,3783 was taken from Netzer et al. (2003)\\
$^{(c)}$ The bolometric luminosity, $L\simeq 9\times \lambda
L_{\lambda}$ (e.g., Kaspi et al. 2000 and references therein) and the kinetic luminosity was calculated in two ways: taking the velocity of the \ion{Si}{14}$\lambda 6.148$ line, and taking the highest velocity measured for any line (for a specific flow component in MCG-6-30-15).}
\end{center}
\end{table*}

\subsection{NGC\,4151}

\begin{figure}
\plotone{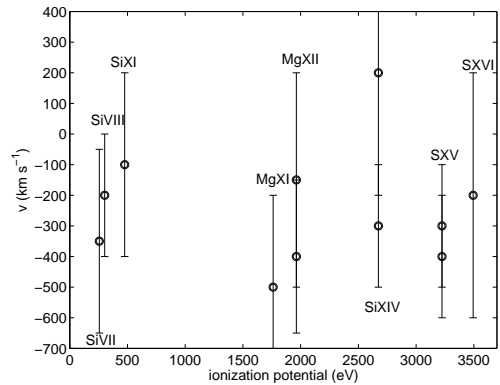}
\caption{Line velocity as measured from {\it Chandra} HETG
observations of NGC\,4151. Here too, our measurements show no clear
correlation of the velocity with ionization potential.}
\label{fig9}
\end{figure}

\begin{figure}
\plotone{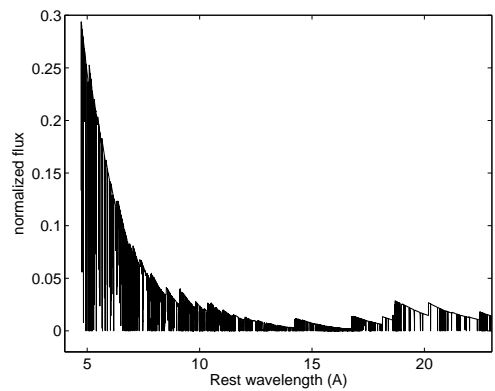}
\caption{The best-fit model for the transmitted spectra through the
outflowing for NGC\,4151 shows considerable opacity at long wavelengths. Clearly, absorption is also important
even at short wavelengths (cf. Kraemer et al. 2005). Accounting
for the absorption we obtain a canonical photon powerlaw index, $\Gamma_{\rm
ph}=1.6$ (see text).} \label{fig10}
\end{figure}

\begin{figure*}
\plotone{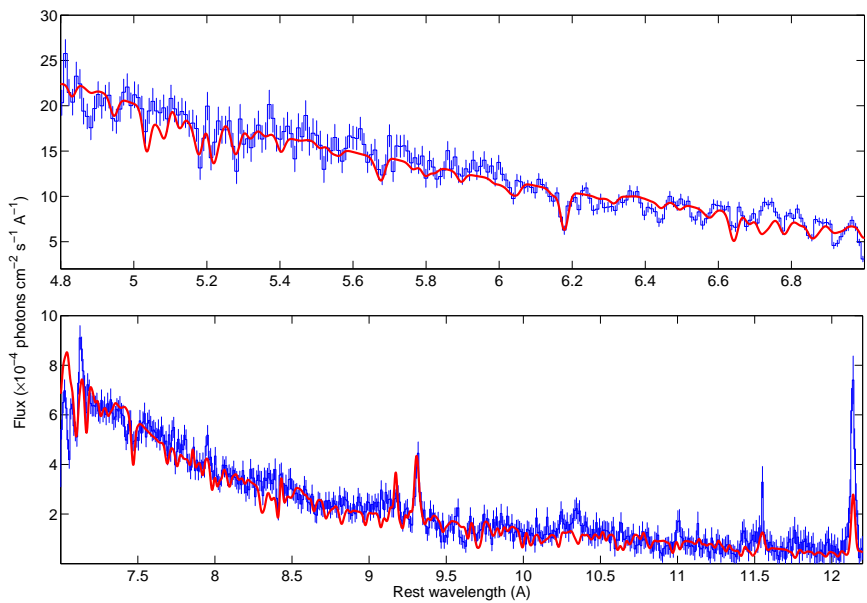}
\caption{An outflow model overlaid on the {\it Chandra/HETG} spectra
of NGC\,4151 (corrected for galactic absorption with a column
$2.2\times 10^{20}~{\rm cm^{-2}}$; George et al. 1998). Clearly,
the fit is able to reproduce many spectral features including 
absorption lines and the shape of the ionizing continuum. That said, the model
fails to account for all features in the spectrum and under-estimates the strength of some emission lines (see text).}
\label{fig11}
\end{figure*}

NGC\,4151 is classified as a Seyfert 1.5. Its UV spectrum is
characterized by multi-component absorption systems with a range
of outflow velocities (0-1600\,${\rm km~s^{-1}}$) and with derived
absorber distances in the range 0.03-2100\,pc (Kraemer et al.
2001). The object is known to be highly variable in the X-rays
(cf. Perola et al. 1982, 1986) showing a complex spectral behavior
indicating possible transverse motion of the X-ray absorber and/or
ionization changes triggered  by flux variation (e.g., George et
al. 1998). Ogle et al. (2000) have resolved extended emission on
kpc scales which motivated Schurch et al (2004) to apply an
ionization cone model while trying to establish an X-ray
unification scheme for AGN. The HETG spectra (see also Kraemer et
al. 2005) is emission dominated for wavelengths greater than $\sim
9\AA$ and the continuum is hard. Luckily, the silicon absorption
line band at short wavelengths has good S/N which allows a
detailed investigation of the HIG flow. We have analyzed all the
HETG observations for this object. In the two more recent
observations (ObsID 3052 \& ObsID 3480) the X-ray flux and the
spectral shape were very similar while in a previous one (ObsID
335), the hard X-ray continuum level was lower and the spectral
shape different. Thus, we concentrate here on the  two recent HETG
observations and analyze their combined spectra resulting in high
S/N data. Inspection of individual lines reveals that there is no
significant correlation between velocity and ionization level
(figure \ref{fig9}). Also, the high velocity ($\gtrsim 1000~{\rm
km~s^{-1}}$) UV components (component A and B in Kraemer et al.
2001) may have little X-ray opacity and are not clearly
detected here.

Our objective here is to find the simplest model possible (i.e., with
the least number of free parameters) which would explain the
global spectral shape of the source as well as individual line
profiles. Thus, we attempted to apply a single outflow model to the
data {\it assuming} that the underlying continuum shape is similar
to that of NGC\,3783 and NGC\,5548 and that the current continuum
level represents the long term (compared to equilibration and
light crossing timescales) average of the source. We find that it
is possible to meet many of the observational constraints by a
single flow model whose spectral features are shown in figure
\ref{fig10}. Clearly, the model is very optically thick with
considerable absorption even at short wavelengths. The EW of the
narrow iron line ($\sim17.5\pm4$\,m\AA according to our
measurements) is typical of non-absorbed type-I AGN (cf. Kaspi et
al. 2001) and is consistent with a scenario in which the absorber
covers the iron $K\alpha$ line emitting region. As before, we have tried
this model with and without additional heating source. Including
such a source results in a slightly better fit to the lines.  

A comparison between the data and the model is shown in figure
\ref{fig11}. The overall agreement is satisfactory yet there are
notable exceptions such as the under-prediction of some emission
lines even for a flow fully covering the ionizing source. As
noted previously by several authors (e.g., Ogle et al. 2001),
NGC\,4151 shows evidence for extended emission, thus it is not at
all surprising that our model under-predicts some emission lines. 
Best fit model parameters are given in table 2. We note that the 
minimum ionization paramter at $r_0$ is  lower by a factor of a few 
than that which characterizes the other objects in our sample. This 
is, however, consistent with the general picture whereby  our line-of-sight 
through the wind penetrates deeper into parts of the outflow below the 
critical point. At such sections of the outflow, the densities are higher 
and the continuity condition results in the flow having somewhat lower 
ionization levels. 

Attributing the changes in the X-ray spectrum over a period of two
years to line-of-sight crossing of the absorber have led Kraemer
et al. (2005) to conclude that a considerable amount of optically
thick gas lies approximately $10^{17}~{\rm cm}$ from the ionizing
source. This is in rough agreement with our distance estimate for
the optically thick part of the flow (see table 2). Thus, if our model is correct
then the prominent X-ray absorber is probably related to the $D+E$
UV components (Kraemer et al. 2005). A more detailed investigation of the 
UV--X-ray connection in this and other objects is beyond the scope of this paper.

\subsection{NGC\,7469}

\begin{figure}
\plotone{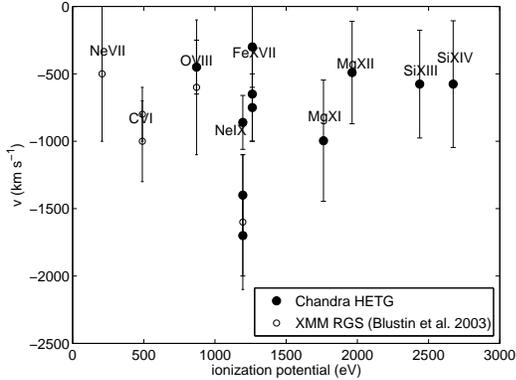}
\caption{Line velocity vs. the ionization potential of the ion responsible for the transition  as measured from {\it Chandra} spectrum of NGC\,7469. The data are augmented by the measurements by Blustin et al. (2003) from
XMM-RGS observation of the object. Clearly, there is no clear
trend between the ionization potential and the line velocity; a behavior which is similar to that seen in other objects in our sample.} \label{fig13}
\end{figure}

\begin{figure*}
\plotone{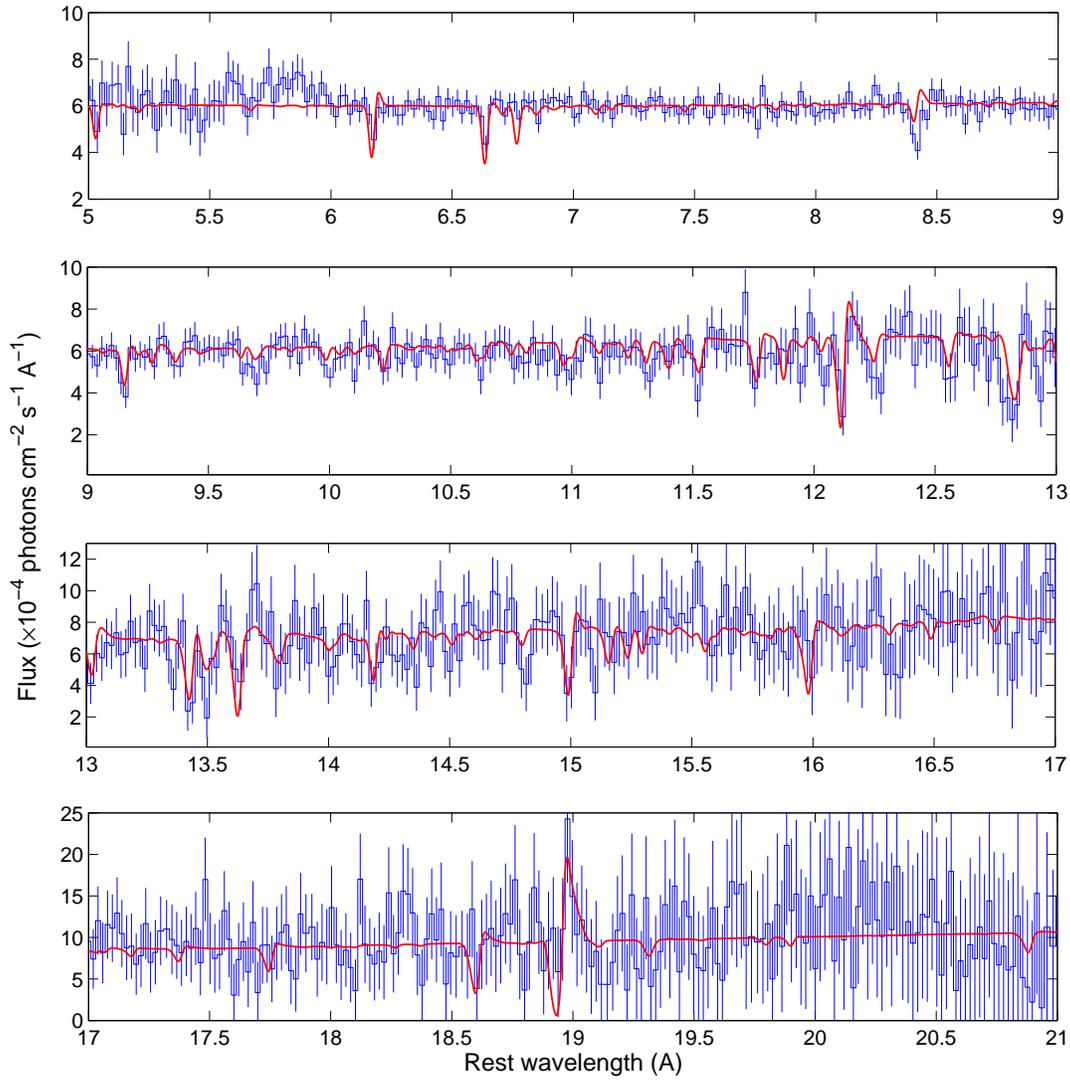}
\caption{An outflow model to the {\it Chandra/HETG} spectra
of NGC\,7469. Clearly, the model reproduces most absorption features. Unlike other objects in our sample, NGC\,7469 shows a soft excess requiring the use of different photon slopes at the short and long wavelength bands (see text). This is not unexpected given that this objects is thought to radiate close to its Eddington rate and, as such, may be similar to narrow line Seyfert 1 galaxies.} \label{fig14}
\end{figure*}

NGC\,7469 is classified as a Seyfert 1.2 galaxy and its mean
luminosity and black hole mass (as deduced from both reverberation
and variability; see Nikolajuk et al.2004) indicate that it is
accreting near the Eddington rate (e.g., Petrucci et al. 2004).
There is evidence for a complex behavior of the X-ray slope, or
alternatively the hardness ratio, as a function of the hard X-ray
flux which is reminiscent of what is observed in NGC\,3783 and
NGC\,5548 (see figure 3 in Nandra et al. 2000). Analysis of the
high resolution X-ray grating spectrum taken for this object by
{\it chandra} shows no significant correlation between the
ionization threshold of the ion and the corresponding velocity
(figure \ref{fig13}).

\begin{figure}
\plotone{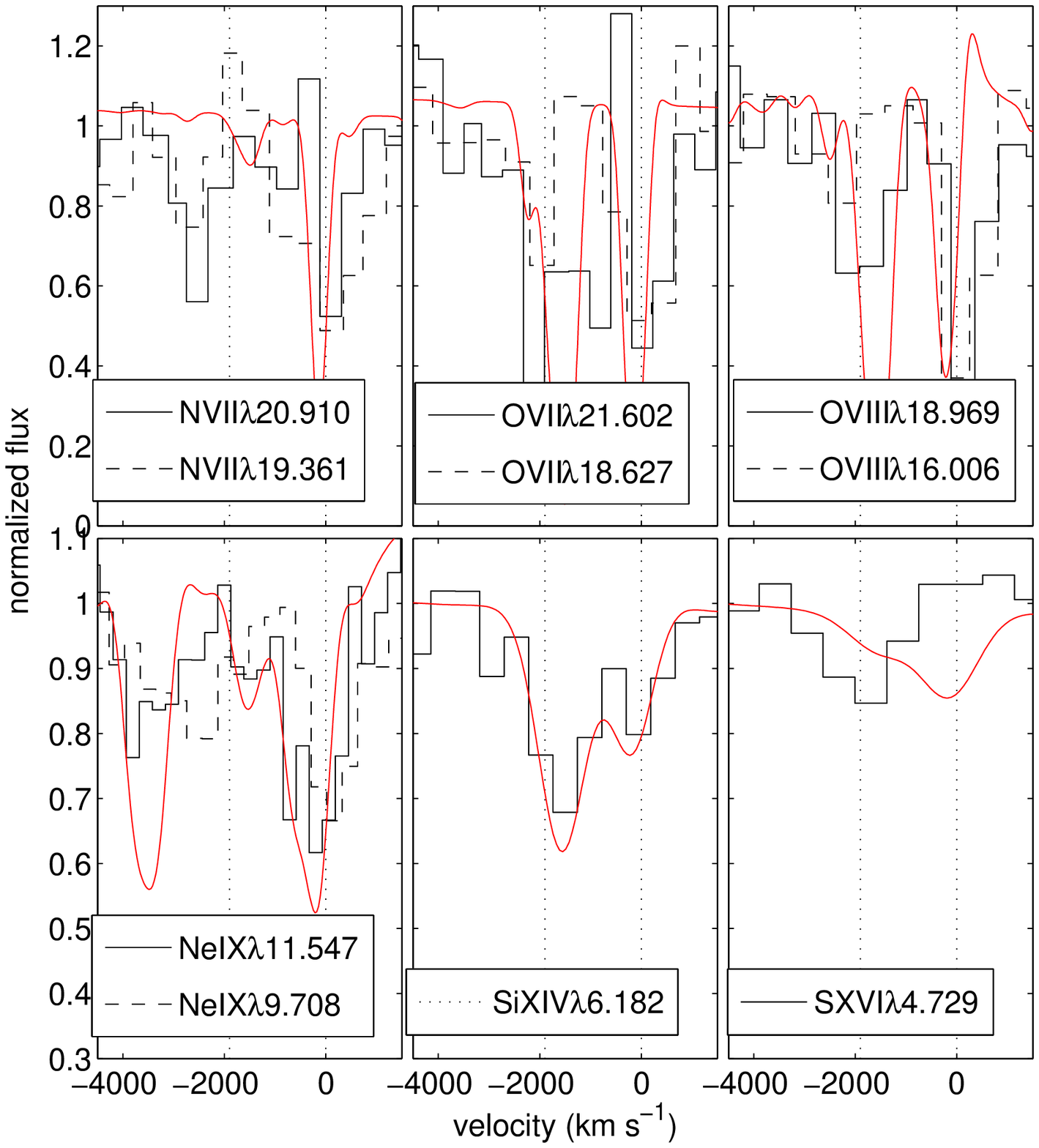}
\caption{Individual line profiles observed in MCG-6-30-15. There
is clear evidence for a two component absorption in the spectrum
with dotted lines serving as guide lines to their velocities (Sako
et al. 2001). Also, more highly ionized lines tend to have a more
substantial optical depth at the high velocity end. The red curve 
shows the specific line profile predictions of our spectral model to the broad-band
X-ray spectrum of this source  (see text and figure \ref{fig17}).} \label{fig15}
\end{figure}

\begin{figure}
\plotone{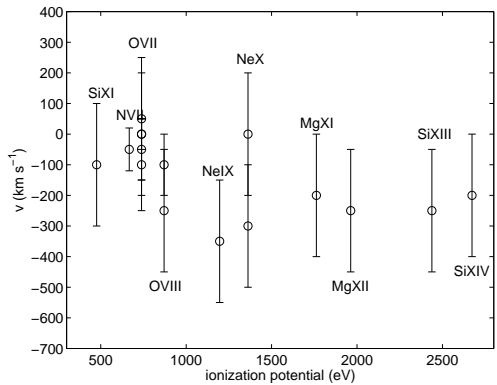}
\caption{Line velocity vs. ionization potential as measured from {\it Chandra} spectrum of MCG-6-30-15 for
the low velocity component of the flow (the high velocity
component is not identified in all lines and is not shown). As in all other cases, there
is no statistically significant trend detected between the quantities.} \label{fig16}
\end{figure}

\begin{figure}
\plotone{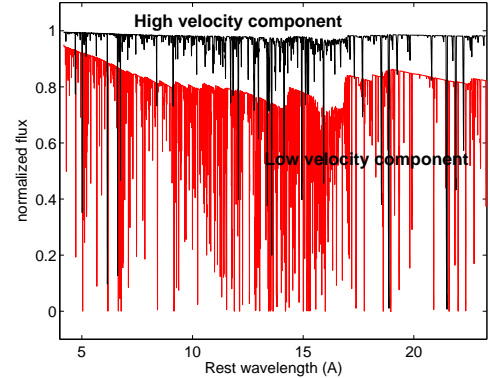}
\caption{The contribution of the low (red curve) and high (black
curve) velocity flows to the opacity of the HIG. Note the low opacity of the
high velocity component relative to the low velocity component
which leads us to conclude that  location estimates based on
column density considerations alone could be misleading (see text).}
\label{fig17}
\end{figure}

\begin{figure*}
\plotone{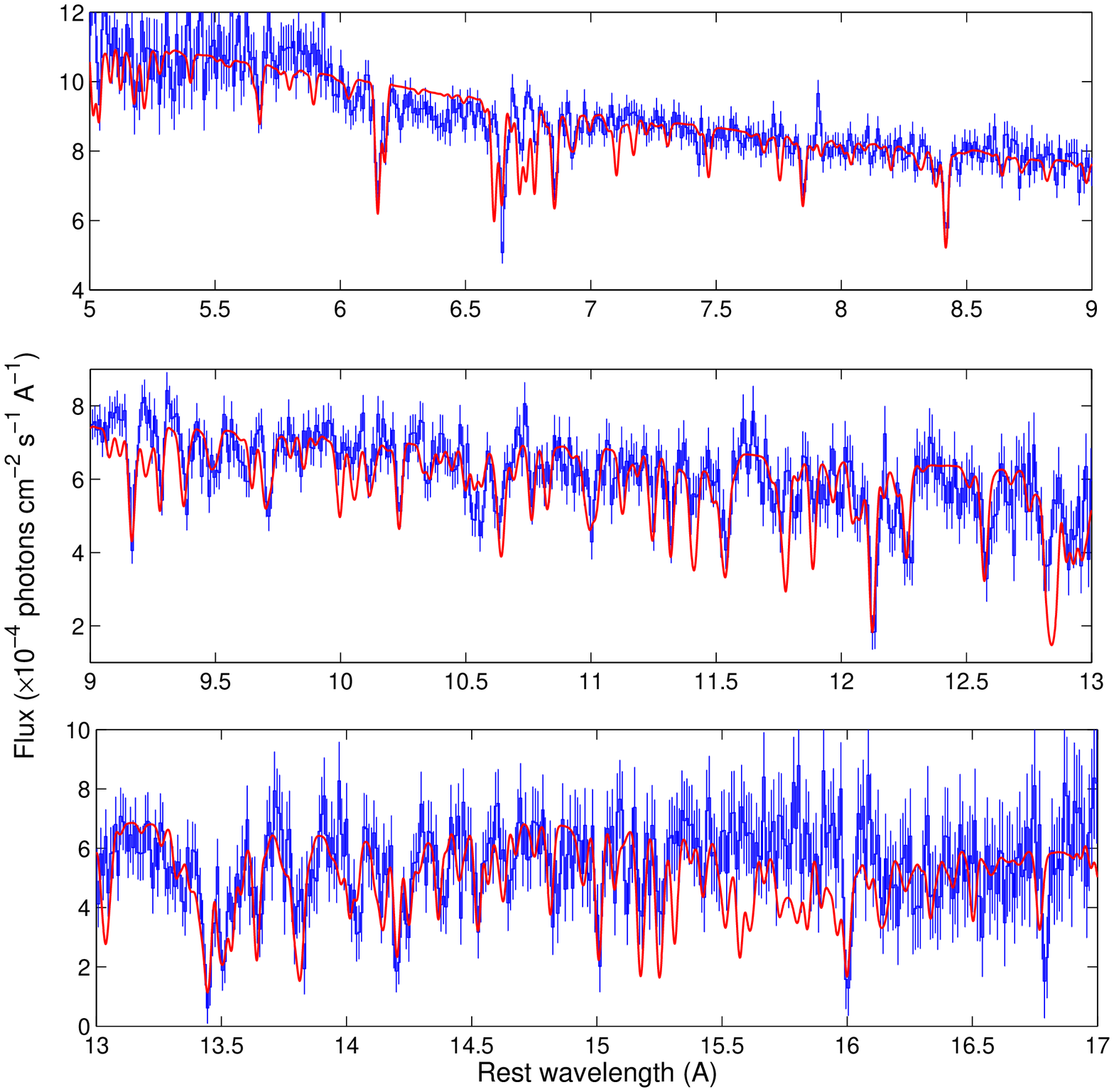}
\caption{An HIG outflow model overlaid on the  {\it Chandra/HETG} spectra
of MCG-6-30-15. Clearly, the model reproduces most of the absorption line features as well
as the overall shape of the continuum.  Some residuals near 15\AA are
likely due to uncertainties in the di-electronic recombination
coefficients of iron (e.g., Netzer 2004).}
\label{fig18}
\end{figure*}

In this case the contribution of radiation pressure force relative
to gravity cannot be neglected since the Compton radiation
pressure force is of the order of the gravitational term. This
results in the effective black hole mass, hence the critical
radius, being smaller. Nevertheless, below the critical point
gravity must prevail for the flow to be stationary. This can
happen if the gas near its footpoint is exposed to a different
ionizing continuum and/or luminosity either by self-shielding,
geometrical effects, or their combination. Another possibility is
that the present continuum level is not representative of its long
term (comparable to or larger than the dynamical timescale) average and
that the flow was launched when the object was emitting at
sub-Eddington rates. Currently, it is not known which of these
scenarios is more realistic and we proceed under the assumption
that the current flux level represents a long term average for
this source.

Given the large uncertainty on the mass and the fact that
radiation pressure force and gravity are of the same order of
magnitude, the uncertainty in the effective gravity is very large.
As the line troughs are not significantly detached the flow
crosses our line-of-sight around its critical point. As there is
no significant soft X-ray absorption, the flow is optically thin
and therefore it is unlikely that gravity prevails at the
sub-sonic part due to considerable opacity. Thus, we continue
under the assumption that the object emits at sub- (although close
to) the Eddington rate. As a starting point we assume it emits at
$0.9L_{\rm Edd}$ and check the sensitivity of the derived flow
parameters to this assumption later on. The effect of adiabatic
cooling is negligible due to the high Eddington rate (Begelman, McKee, \& Shields 1983) 
and  an additional heating source is not required by the data.

We have attempted to explain the continuum shape and lines using a single
powerlaw for the underlying (unabsorbed) continuum extending from soft to hard X-ray energies. Our
calculations show that a single powerlaw model fails to account
for the spectral shape of both the soft and hard X-ray spectral
bands and that the powerlaw steepens towards lower energies. (An
alternative explanation of a single powerlaw with a photon slope, $\Gamma_{\rm
ph}\sim 2.3$ fails to account for the spectral shape near $\sim
1$\,keV and over-predicts the strength of the absorption lines.)
The presence of a soft excess is by no means surprising; in fact,
it is expected to be more prominent in objects accreting near their  Eddington
rate (e.g., Wang \& Netzer 2003, Matsumoto et al. 2004). Our model suggests that a good agreement between the model and the data is obtained if  $\Gamma_{\rm ph}$ for the hard ($>1$\,kev) X-ray band is
$\sim 1.9$ while at lower energies $\Gamma_{\rm ph}\sim 3.5$ (the
relative normalization of which is $\sim0.3$ at 1\,keV). The data
quality is not good enough to tell whether partial covering
effects are important in the soft X-rays since the lines are weak
and the S/N low. As for the emission lines, these are weak and are
likely to be absorbed by the outflowing gas given the low velocity
dispersion of the flow. The data are consistent with an
absorbing/emitting outflow that fully covers the ionizing source
though we  note that this is only an upper limit and that the true
coverage can be much less than unity.

The outflow model is relatively optically thin at all
wavebands and a comparison between the data and the model is shown
in figure \ref{fig14}. The overall agreement is good with no
apparent deviations apart from a calibration issue
shortward of $6\AA$ at the interface between two chips.

The best-fit model parameters are shown in table 2.  We
note that our choice of effective black hole mass was somewhat arbitrary 
and that the source can emit even closer to its
Eddington rate, say at 95\% of its Eddington luminosity. In this
case the effective gravity is smaller and the critical point moves
closer in. Thus, our predicted location of the flow in this object
is rather uncertain. This has some bearing on the other parameters
of the model whose errors are somewhat larger than those for the other
objects in our sample. These uncertainties can only be reduced by
better measuring the black hole mass and determining the shape of
the ionizing continuum, or by measuring ionization equilibration timescales
for the plasma. It is interesting to note that, given the
uncertainty on the effective gravitational term, the flow can, in
principal, be launched just outside the broad line region (BLR)
for this object (located at around $10^{16}~{\rm cm}$; Kaspi et
al. 2000). This, however, requires somewhat different parameter values than those derived for flows
in other objects. Specifically, for flows which are launched much
closer to the inner engine, larger values of $\beta$ are needed to maintain the same opacity as the flow is launched closer to the ionizing source.
We find this explanation less likely since $\beta$ is expected to
be governed by the microphysics of the gas rather than the
properties of the AGN. Thus, based on the similarity of flows in
different objects we expect the flow in NGC\,4151 to be launched
at somewhat larger distances than the BLR.

\subsection{MCG-6-30-15}

MCG-6-30-15 is a borderline narrow Seyfert 1 galaxy that has been
the center of an ongoing debate concerning the properties of
its soft X-ray spectrum (see Sako et al. 2001). We wish to emphasize that here we do not attempt to resolve this debate which, in our opinion, requires a deeper understanding of the soft X-ray spectrum of AGN. While in the {\it ASCA} days the underlying ionizing continuum of AGN was often modeled by a single powerlaw,  accumulating {\it Chandra} and {\it XMM} observations have shown that this is not the case (e.g., Netzer et al. 2003, Page et al. 2004). For NGC\,3783 a soft excess is clearly visible despite the complicated soft X-ray absorption. Other objects, for which absorption is less important, show that, at least in some cases, there is strong evidence for an X-ray hump at soft energies (e.g., Turner et al. 2001, Ogle et al. 2004, Vaughan et al. 2004). The underlying physical mechanism for the soft X-ray hump or excess is unknown and may be related to the accretion disk (e.g., Wang \& Netzer 2003), relativistic soft X-ray emission lines (e.g., Branduardi-Raymont et al. 2001), a blend of soft X-ray emission lines (e.g., Pounds et al. 2005), or relativistically smeared absorption (Gierli{\'n}ski \& Done 2004). While all of the above applies to some extent to all AGN, the case of MCG-6-30-15 seems to be unique in the sense that the soft X-ray spectrum is poorly understood, with recent works attempting to explain it as either due to relativistic line emission (e.g., Sako et al. 2003) or dust-related absorption features (e.g., Lee et al. 2001, Ballantyne et al. 2003). 

Given the aforementioned uncertainties concerning the X-ray continuum shape at soft X-ray energies we concentrate in this work on the hard X-ray continuum only and note that a well constrained solution may be obtained by fitting for the silicon and sulphur inner shell lines near 6\AA (CN05, N03). For this purpose we have analyzed the recent {\it Chandra}/HETG observations and confirm the existence of
two kinematical components (Sako et al. 2003) that are shown in figure \ref{fig15}.
Like in NGC\,3783, there is no clear correlation
between the ionization level and the velocity of the slower component of the flow (figure
\ref{fig16}). Furthermore, the hint for higher ionization levels having somewhat higher outflow velocities is evident from figure \ref{fig15} where the slower kinematical component has less opacity at high ionization levels. We note that not all lines show clear signature for a high velocity component
which is best visible in the \ion{O}{7}, \ion{O}{8}, and \ion{Si}{14} lines. In
fact, some lines do not show the high velocity component as is
evident from figures \ref{fig1} \& \ref{fig15}. We note that high
ionization lines (e.g., \ion{Si}{14} and \ion{S}{16} tend to have
stronger absorption at higher velocities and are consistent with
being related to the high velocity component. Judging by the relative strength of higher order resonance transitions it seems that the high velocity flow component has an overall lower column density.
We have searched for the existence of the two component flow in
the UV band but the poor quality of the IUE data precludes a
definite answer to this question (e.g., Reynolds et al. 1997).

It turns out that the physics of the two component outflow cannot
be easily explained in the framework of a single outflow model discussed here.
One of the problems relates to the lack of significant absorption
by high ionization lines at low velocities (e.g., the
\ion{S}{16}$\lambda4.729$; see figure \ref{fig15}). As discussed
here and in CN05, lines coming from the same flow are likely to
have very similar line profiles, this was actually the motivation
behind the model for the outflow in NGC\,3783 (CN05). To reach a velocity of
order $2000~{\rm km~s^{-1}}$ in thermally driven isothermal flows,
the sonic velocity should be roughly $400~{\rm km~s^{-1}}$; i.e.,
corresponding to highly ionized gas at very high temperatures. Thus, if the low velocity component is part of the high velocity flow then it lies in the sub-sonic region. Such region is likely to be very optically thick (see \S4.2) which is not the case here. In addition, to explain the velocity range devoid of any
absorption, it is required that the gas be transparent at those
velocities; either due to the lack of relatively cold, hence
observable gas or due to the lack of gas altogether. In the
latter case, the flow cannot be considered as a single entity and
the simplified equation of motion used here does not hold. Also, the lack of significant
high ionization absorption across the entire velocity range
occupied by the two components strongly suggests they are indeed
distinct physical components.

It is worth noting that if, as suggested by some authors, the outflowing gas is dusty then its kinematics may be very different than assumed here (since dust particles have a large cross-section for the absorption and scattering of radiation) and could alter our results (e.g., Everett 2002). That said, the lack of conclusive evidence for the existence of dust in this object and in other objects in our sample, lead us to neglect its effect in the current study. This supports the claim by Ballantyne et al. (2003) stating that dust, if present, is likely to be associated with much larger scales.

In what follows we therefore assume that the two components belong to
two different flows or streams and model those accordingly. We first fit the low velocity components which seem to be
responsible for most of the X-ray opacity (see above), and then
add the high velocity flow component. We assume that both components of the outflow
completely cover the ionizing source. The two component
model is shown in figure \ref{fig17} and the model parameters
given in table 2. We note that the high velocity flow is
relatively optically thin and therefore continuum shielding by it is negligible. The high velocity model
requires an additional heating source apart from photoionization
to overcome the effects of adiabatic cooling and drive the flow to the observed
velocities. This is not the case for the low velocity
component where equally good fits are obtained in either case
(i.e., with and without additional heating source). Our best fit
model is shown in figure \ref{fig18} for which an underlying canonical photon
powerlaw of $1.6$ has been assumed. While the model does have
considerable deviations near $16\AA$ (probably due to the atomic uncertainties
discussed above), the overall agreement is satisfactory.  A zoom in on individual lines is shown in
figure \ref{fig15} where the model provides a reasonable fit for
most lines with the exception of \ion{S}{16}$\,\lambda 4.729$ for
which it over-estimates the opacity at the low velocity end. 

We emphasize again that we have made no attempt to model the soft X-ray spectrum of this object, despite the good S/N, since our understanding of the various soft X-ray emission components in AGN is limited with MCG-6-30-15 being perhaps a more extreme manifestation of that problem.

While the flows are not co-spatial (the higher velocity stream is launched from roughly an order of magnitude closer in to the ionizing source) interaction between the different streams might occur. The detailed modeling of which is beyond the scope of this paper and requires detailed numerical calculations combining detailed atomic physics and hydrodynamic calculations.

\section{Discussion}

\begin{figure*}
\plotone{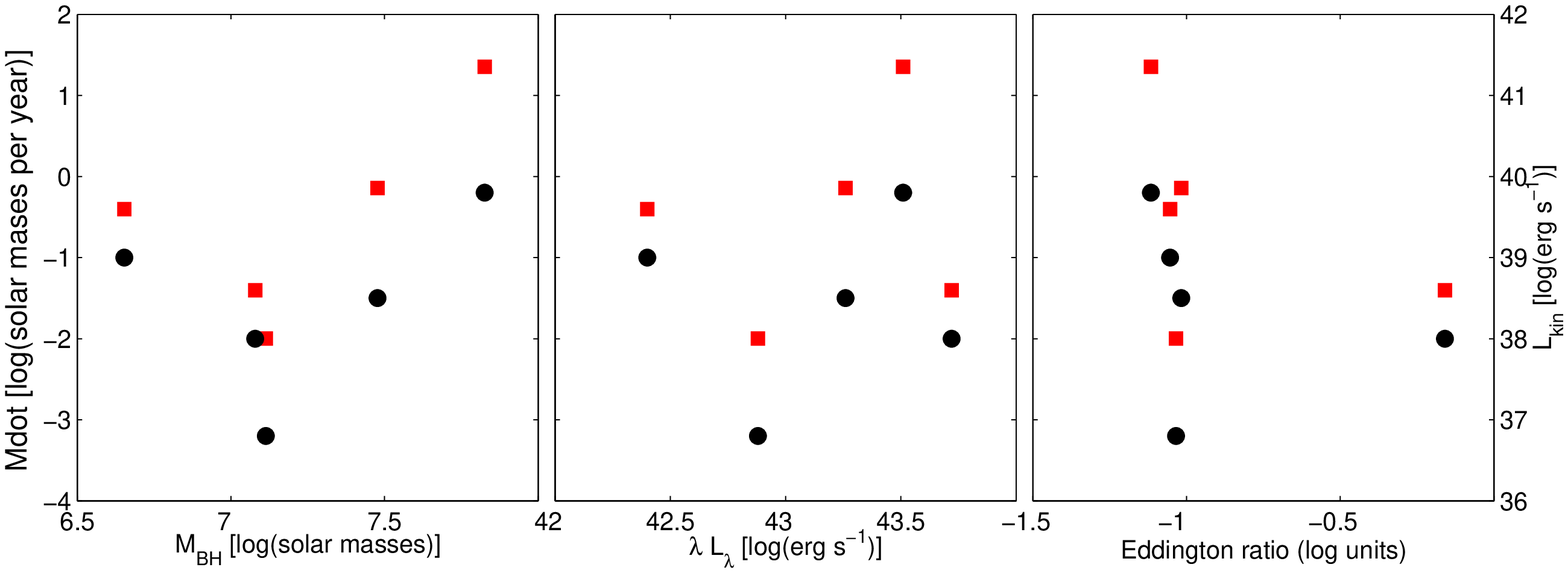}
\caption{Derived mass loss rates (black symbols) and kinetic
luminosities (red symbols) as a function of the black hole mass
(left panel), optical luminosity (middle panel), and the estimated
Eddington ratio (right panel). The kinetic luminosity was
calculated using the average velocity of the unblended lines shown in figures 7, 10, 13, \& 16 (for NGC\,3783 we have used the Kaspi et al. (2002) measurements). There is a hint for a trend where more massive and more luminous objects tend to have more massive and more energetic outflows. We do caution the reader than the mass loss rate has a typical uncertainty of 0.5\,dex (not shown, see table 2). In addition, objects with similar emission efficiencies can have very different mass loss rates and kinetic luminosities.} \label{fig19}
\end{figure*}

The purpose of this paper is to determine the mass loss rate and kinetic luminosities associated with outflows in AGN.  While similar studies have been possible for stars in which many absorption lines are visible against the ionizing continuum, in AGN there are much fewer lines (essentially none in the optical and only handful in the UV) and the ionizing continuum extends from the UV to the X-rays where many resonance lines are present. For this reason, high resolution, good S/N X-ray spectroscopy is crucial for understanding the dynamics of such flows in those objects.

The main advantage of the model presented here over other methods in the literature is its ability to provide a more physical rather than phenomenological understanding of photoionized flows. It is not the purpose of this paper to provide the best possible fit to the data by a $\chi^2$ standard. While such a method has proven adequate for low resolution CCD spectra, it was shown to be inadequate to model the high-resolution, rich spectra of AGN (N03). Furthermore, obtaining good fits to the data has been successfully done by using pure photoionization models. In particular, for the case of NGC\,3783 a three-zone model has been suggested by N03 with 12 free parameters (a column density, ionization parameter, global covering factor of the ionizing source, and an outflow velocity for each zone). While providing a good fit to the absorption spectrum, this method cannot be used to determine the density or distance of the gas from the ionizing source (unless variability arguments are used) and the mass loss rate and kinetic luminosities are essentially undetermined. Contrary to that, only five parameters are used in this work ($U_{\rm ox}^{\rm min},~U_{\rm ox}^{\rm max},~\beta,~r_0$ and $C_{\rm global}$) that determine not only the thermal and ionization properties of the flow but also its kinematic properties revealing its location, mass loss rate, and kinetic luminosity.

All AGN in our sample show evidence
for very highly ionized gas outflowing from their centers. Such gas would have
eluded detection by previous X-ray missions and suggests that either such outflows
are common in AGN or that they are associated with much cooler gas (e.g., the warm absorber) in those 
objects. All flows are consistent with a picture in which the gas is  multi-phased with higher ionization 
more dilute phases filling most of the volume. Our best-fit model parameters for flows in
different objects seem to converge to similar values  implying similar physics for the HIG in different
objects. The flows studied here are thermally driven with radiation pressure force being negligible but in objects shining close to their Eddington luminosity.

Our study shows that, despite the very different manifestations of
HIG flows in AGN (e.g., some being more optically thick or having
higher velocities than others), the microphysics of all flows is
remarkably  similar. For example, the wind is multiphase spanning some three orders of 
magnitude in density at any location with a similar density-scale powerlaw dependence 
(i.e., similar $\beta$). Our analysis also suggests that in objects that shine much below their Eddington rate, and whose wind velocity exceeds $\sim 1000~{\rm km~s^{-1}}$, an additional heating source is probably required to balance the effect of adiabatic cooling. (There are several possible such heating mechanism among which is by acoustic waves which are emitted by the cool phase of the gas as it reacts to flux variations of the ionizing source; see Chelouche 2007.)  This uniformity of  the outflow properties between very different objects 
is not unexpected since the microphysics of the gas is unlikely to be considerably affected by the macro-physics of the AGN. In fact, this adds credibility to the model in
the sense that consistent results are obtained for all objects despite their very different spectra. Furthermore, the value for
the ionization parameter at the critical point (derived from table
2) seems to be similar for most objects in our sample. In cases
where the flow is more optically thick (e.g., in NGC\,4151), the
measured velocity is low and our model suggests we may be looking
deeper towards the footpoint of the flow. Thus, if our model is correct then 
it seems that the geometry of the outflowing gas plays a role  in the spectral manifestation of type-I objects. This conjecture, however, is based on a handful of objects and is yet to be
confirmed observationally and better modeled, theoretically (we note that for
thermal driven flows, the notion of stream bending
as is the case for radiation pressure driven flows; e.g., Pereyra
et al. 1997).

Our results concerning the launching radius of the flow - which we
identify with $r_c\sim 1$\,pc - is much larger than the size of the inner
accretion disk in those objects. This disagrees with some versions 
of the AGN unification scheme, which suggest that such flows are launched 
from very close to the black hole in those systems (e.g., Elvis 2000). As
discussed in CN05, a promising mass reservoir for these outflows is the putative torus that 
is thought to lie on scales similar to $r_c$. Alternative explanations may involve
association with the central star cluster. For one object in our
sample (NGC\,4151; see \S4.2) it seems that the outflowing gas
obscures  the emission region of the iron ${\rm K\alpha}$ line
whose EW is similar to that of typical, non-obscured, type-I sources.
Thus, the HIG absorber seems to lie farther out beyond the iron
$K\alpha$ emitting region, which is usually identified
with the torus (e.g., Netzer et al. 2002). This leads us to
conclude that absorption by highly ionized outflows is unlikely to
be the cause for obscuration in type-II AGN.

The results regarding the emission from such winds indicate that
the global covering fraction is consistent with being of order unity but can be much smaller than that if part of the emission comes from different regions not associated with the flow (as is the case for NGC\,4151; Ogle et al. 2004). An extensive study of
the possible contributors to the emission spectrum in type-I and
type-II objects is beyond the scope of this paper.

The deduced mass loss rates from AGN indicate that they are roughly of the order of 
the mass accretion rate yet with a large scatter (see table 2). Thus, to maintain persistent
accretion over the AGN lifetime, a large gas reservoir is needed
or else a large deposition of matter into the central regions of
the galaxy. As the flow velocities are much smaller than the speed of light, the associated kinetic luminosities
are very small (typically much less than 0.1\%) compared to the bolometric one. The large scatter in the deduced mass loss rate and kinetic luminosity can be traced to the typical velocity of the outflow which varies by a factor of several among objects. This large scatter may be reduced if the true outflow velocity is similar in different objects while the observed velocity is subject to inclination effects. While theoretically plausible, there is no observational support for such a claim and we do not discuss it any further in this work.

Given our results for the mass loss rate and kinetic luminosity of
flows we have searched for trends between those quantities and other
properties of the AGN. These are shown in figure \ref{fig19}. We do not find any  significant correlation between the mass loss rate,
$\dot{M}$, or the kinetic luminosity, $L_{\rm kin}$ and the object's
luminosity, black-hole mass, or the Eddington ratio. There may be
a hint for a trend between $\dot{M}$ and $L_{\rm kin}$
and the black hole mass and luminosity which may be driven, in part, by the possible trend between black hole mass and flow velocity (see figure 1).

Extending the diagrams beyond the narrow luminosity range reported
here requires detailed observations of more distant quasars as
well as low luminosity AGN. It is not at all clear what the
spectral signatures from such HIG flows, if at all present, would
be in those objects. Nevertheless, for the sake of (perhaps unjustified)
simplicity, let us assume that the general properties of the model hold also at higher luminosities and more massive objects. Let us further assume that we increase the object luminosity, $L$, and black hole mass, $M_{\rm BH}$ in the same proportion so as to maintain a fixed $L/L_{\rm Edd}$. In this case, for a fixed ionization paramter, the density would scale as $L/M_{\rm BH}^2$. Thus, the column density $n\dot r \propto L/M_{\rm BH} \propto L/L_{\rm Edd}={\rm const.}$ and similar gas opacity, hence absorption features, are expected to occur also in more and less luminous objects for a fixed Eddington rate. The mass loss rate as well as the kinetic luminosity would then be proportional to $M_{\rm BH}$ (or $L$). Thus, if our model can be extended to high luminosity objects and such objects emit at the same Eddington ratio as Seyfert galaxies one naivly expect the kinetic luminosity to be a constant fraction of the bolometric luminosity (i.e., much less than one per cent). Di Matteo et
al. (2005) assumed in their simulations that $L_{\rm kin}/L\simeq
5\%$, i.e., much larger than the estimates given here.  Thus, if our conclusions for the relative kinetic luminosity hold also for highly luminous quasars then such flows cannot account for the necessary feedback effect.

Lastly we note that our spectral fits point to a
canonical X-ray spectrum for AGN accreting at sub-Eddington rates
in which the hard photon X-ray slope is $\sim 1.6$ and that any
apparent changes to this slope are due to absorption effects. In
the soft X-rays there is evidence for a soft excess in some but
not all objects, the physical nature of which is currently unclear.
The spectral slope is more steep for the one object in our sample which seems to accrete near its Eddington rate. We caution, however, that despite the high quality spectrum and the detailed analysis presented here these conclusions are based on a handful of objects and a more detailed investigation of the complex soft X-ray spectrum is needed to address these issues.

\section{Conclusion \& Summary}

We present the first systematic study of  type-I AGN with high resolution spectroscopic 
X-ray observations. Our
sample consists of five objects for which good signal-to-noise  {\it Chandra/HETG} data
 are available. We report on our findings concerning the physical properties of
highly ionized, low velocity gas seen to be outflowing from those systems.  Using a novel scheme for the
calculation of the spectral features from thermally and
radiatively driven winds we are able to constrain, for the first time, physical
properties of the ejected gas pertaining to its location, the mass loss rate, and the associated kinetic luminosity. In addition, the number of free parameters in our model is typically much smaller than that used by pure photoionization models. We find that all AGN in our sample possess highly ionized outflows whose mass loss rate is, on average, of
order of the mass accretion (but with a scatter of order a few around this value) while the 
kinetic luminosity is only a small fraction (typically $\ll 1\%$)  of the 
bolometric luminosity. If this
scaling persists to highly luminous quasars then low velocity
flows are unlikely to lead to substantial feedback effects. Our
results indicate that such flows show an interesting physical similarity
among outflows in different objects suggesting that the gas microphysics in those systems is
essentially the same. Different manifestation of such flows can be
attributed, at least in part, to geometrical effects. Furthermore,
we find that these flows are launched from parsec scales  and are 
unlikely to be associated with the inner parts of the accretion disk. 
Our study demonstrates the importance of high resolution and S/N X-ray 
grating observations  for understanding the physics of AGN flows. 

\acknowledgments
We thank S. Kaspi for his invaluable help in dealing with the X-ray spectra of 
NGC\,5548 and H. Netzer for helpful comments.

\end{document}